\documentclass[a4paper,pre,floats,nofootinbib,showpacs,superscriptaddress,floatfix]{revtex4}
\usepackage{epsf,graphicx}
\usepackage{subfigure}
\usepackage{latexsym,amsmath,verbatim}
\usepackage{vmargin}

\setmarginsrb{3cm}
{2cm}
{3cm}
{4cm}
{0mm}{0mm}{0mm}{0mm}

\begin{document}

\title{MINIMAL AGENT BASED MODEL FOR THE ORIGIN AND SELF-ORGANIZATION OF STYLIZED FACTS IN FINANCIAL MARKETS}
\author{V. Alfi}
\affiliation{Centro Studi e Ricerche ``E. Fermi''\\ Compendio del Viminale, 00184, Roma, Italy}
\affiliation{Dipartimento di Fisica, Universit\`a ``La Sapienza''\\P. le A. Moro 2, 00185, Roma, Italy}
\author{L. Pietronero}
\affiliation{Dipartimento di Fisica, Universit\`a ``La Sapienza''\\P. le A. Moro 2, 00185, Roma, Italy}
\affiliation{``Istituto dei Sistemi Complessi'' CNR\\ V. dei Taurini 19, 00185 Rome, Italy}
\author{A. Zaccaria}
\affiliation{Dipartimento di Fisica, Universit\`a ``La Sapienza''\\P. le A. Moro 2, 00184, Roma, Italy}

\begin{abstract}
We introduce a minimal Agent Based Model with two classes of
agents, fundamentalists (stabilizing) and chartists (destabilizing) and we focus on the essential features which can generate the stylized facts. This leads to a detailed understanding of the origin of fat tails and volatility clustering and we propose a
mechanism for the self-organization of the market dynamics in the quasi-critical state.
The stylized facts are shown to correspond to finite size effects
which, however, can be active at different time scales.
This implies that universality cannot be expected in describing these properties in terms of effective critical exponents.
The introduction of a threshold in the agents' action (small price fluctuations lead to no-action) triggers the self-organization towards the quasi-critical state.
Non-stationarity in the number of active agents and in their action plays a fundamental role. The model can be easily generalized to more realistic variants in a systematic way.
\end{abstract}
\maketitle

\section{Introduction}
The Random Walk (RW) represents the simplest modelization for the behavior of financial time series. It was in fact introduced by Louis Bachelier~\cite{bachelier} in 1900 to model the dynamics of prices. Recently, the availability of large amounts of data has been accompanied by a systematic study of the empirical properties of price time series. This led to the identification of relatively well defined properties, the ``stylized facts'' (SF), reasonably common to all markets, which are beyond the simple RW dynamics~\cite{bachelier,rcont,rcont2,lebaron,farmer,mantegna,bouchaud,mandelbrot}.
These properties refer to the time series of the prices $p(t)$ of the return $r(t) = p(t) - p(t-dt)$. 
The main SF are the absence of linear autocorrelation of returns, heavy tails and volatility
clustering.
Non stationarity is also a characteristic property of price dynamics which can lead to
difficulties in the statistical analysis of the data.
The extended correlations implied by the above properties and their
possible power law behavior led many authors
to conjecture a similarity between these properties and critical exponents in statistical physics.
This line of reasoning leads naturally to models with heterogeneous agents which can lead to large fluctuations and emerging properties. In the past years many models of this type have been introduced. However, as noted by LeBaron, Hommes and other authors~\cite{lebaron2,hommes,samani},
often the complexity of these models does not permit to clearly point out which aspect of the model is really responsible for the SF and to which extend these results can be considered as an 
explanation of the empirical evidences.
An additional point which is usually neglected and that we are going to consider in detail is that of self-organization~\cite{perbak,giardina}, namely the fact that the dynamics evolves spontaneously towards
the range of parameters corresponding to the quasi-critical state with the SF. 
A general characteristic of several ABM, as for example the Lux-Marchesi model~\cite{lux1,lux2,lux3},
is the competition between stabilizing and destabilizing agents, corresponding usually to fundamentalists and chartists.
The fundamentalists assign a value to a stock $(p_f)$ which is defined by classical economic arguments. They buy or sell depending on whether the actual price $p(t)$ is below or above the reference value $p_f$. The chartists instead only consider the price time series and are typically trend followers.
The competition between these two classes, together with the possibility of changing strategy,
is believed to be a realistic and generic property of the markets.
Many of these models indeed lead to the SF for same range of parameters but
the identification of their specific origin it is often hidden in the complexity
of the model.
Even the basic problem whether they correspond to genuine critical exponents or finite
size effects,  is an open question~\cite{stauffer}.

In this paper we introduce a minimal model with the objective of reproducing the SF with the simplest possible elements, in order to trace precisely the origin and the nature of the SF.
We find that the SF arise from \underline{finite size effects}, namely
they disappear at infinite time or for an infinite number of agents.
This has deep implications for the conceptual structure of the models as well as for
data analysis because, in such a situation, one cannot expect properties
of universality.
In this perspective the SF arise from the same general phenomenon but their specific realization in different situations may lead to different quantitative properties.
Also the interpretation in terms of critical exponents should be considered more as a fitting scheme rather than as a fundamental characterization.
The possibility to achieve a detailed understanding of the origin of the SF permits then to pose the fundamental problem of their \underline{self-organization}. A crucial point in this respect is the variability of the total number of agents $N$ as a function of price fluctuations. We show that the introduction of a threshold to enter or exit the market may naturally lead to self-organization.
Finally we would like to stress that the general perspective of the present model is to have the maximum simplification in order to identify the specific origin of the SF. The model can be easily generalized to higher degrees of realism in systematic way~\cite{alfi1}.

\section{Simplest Agent Based Model with Fundamentalists and Chartists}

We consider $N$ agents divided in $N_f$ fundamentalists and $N_c$ chartists $(N=N_f+N_c)$. For the moment we keep $N$ fixed but later, in the study of  self-organization, $N$ will be a fluctuating variable. An important point is to have a description of chartists as simple as possible and with the minimal number of parameters. In this respect it is useful to describe the chartists in terms of the effective potential method~\cite{taka1,taka2,alfi2,alfi3} and consider only the trend follower (destabilizing) case.
The signal considered by a chartist is the distance between the price $p(t)$ and a suitable moving average $p_M(t)$ defined as:
\begin{equation}
 p_M(t)=\frac 1M \sum_{k=1}^{M}p(t-k)
\label{eq:1}
\end{equation}
where $M$ is the number of time steps on which the moving average is computed.
The basic idea is that the difference between the actual price $p(t)$ and the its smoothed profile (moving average) represents a signal for the chartists who will bet on the amplification of this difference. This tendency is described in terms of a repulsive force which applies to the dynamics of the price and which is proportional to the distance $p(t)-p_M(t)$. The corresponding potential is:
\begin{equation}
\Phi(p(t)-p_M(t))=\frac{b}{2(M-1)}(p(t)-p_M(t))^2
\label{eq:2}
\end{equation}
where $b$ is the strength of the potential and the term $M-1$ makes the potential independent
on the time window $M$~\cite{taka1,taka2,alfi2,alfi3}.
\begin{figure}[htbp]
\centering
\includegraphics[scale=0.4]{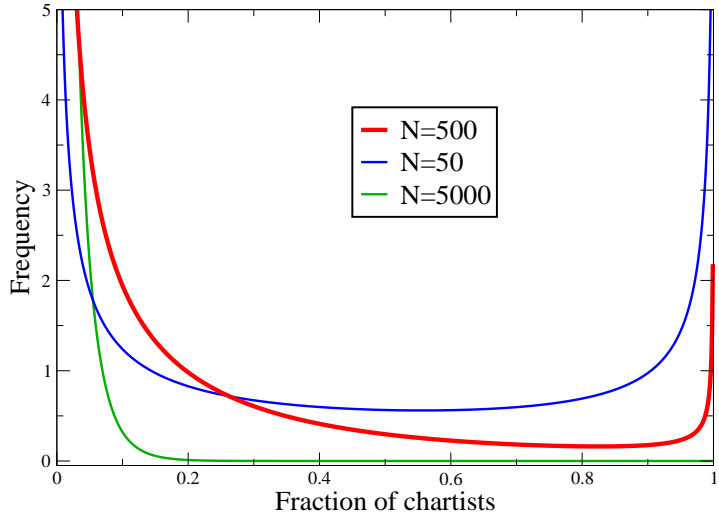}
\caption{Stable asymptotic distributions for the nature of the agents. In the left side fundamentalists dominate while chartists dominate on the right side. The model is slightly asymmetric in favor of fundamentalists. However, for a small number of agents (50) the blue line shows an almost symmetric distribution. For larger values of $N$ (500 and 5000) the transition probability between metastable states decreases and one can see that fundamentalists become dominant.}
\label{fig:1}
\end{figure}
For the fundamentalists the signal is given by $p(t)-p_f$  and the total stochastic equation for the price dynamics is:
\begin{equation}
p(t+1)=p(t)+\sigma\xi+ b \frac{(p(t)-p_M(t))}{M-1} n_c+\gamma(p_f-p(t))n_f
\label{eq:3}
\end{equation}
The term $\sigma\xi$ represents a white noise while $n_c=N_c/N$ and $n_f=N_f/N$ are the relative number
of chartists and fundamentalists. The parameters $b$ and $\gamma$ represent the impacts on the price due to chartists and fundamentalists respectively. The dynamics is defined in terms of unitary time steps and the impact of the agents on the market is proportional to their signal.
This is a specific realization of Walras law~\cite{mantegna,bouchaud,johnson} for the price formation in terms of the excess demand $ED$, which we use here in the simple linearized form.
\begin{equation}
\frac{dp}{dt}= ED
\label{eq:4}
\end{equation}
\begin{figure}[htbp]
\centering
\subfigure[]{
\label{fig:2:a}
\includegraphics[scale=0.4]{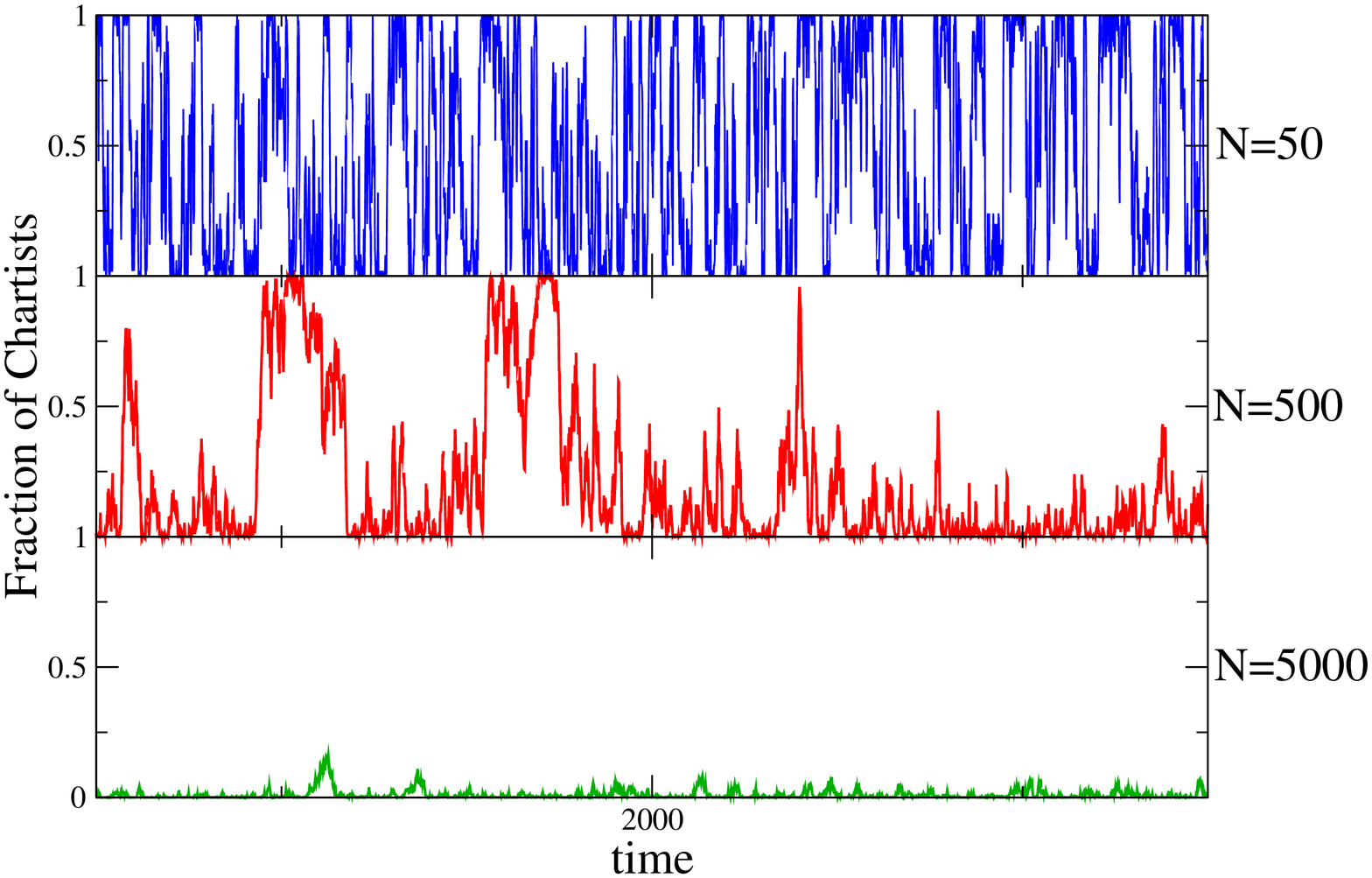}}
\subfigure[]{
\label{fig:2:b}
\includegraphics[scale=0.4]{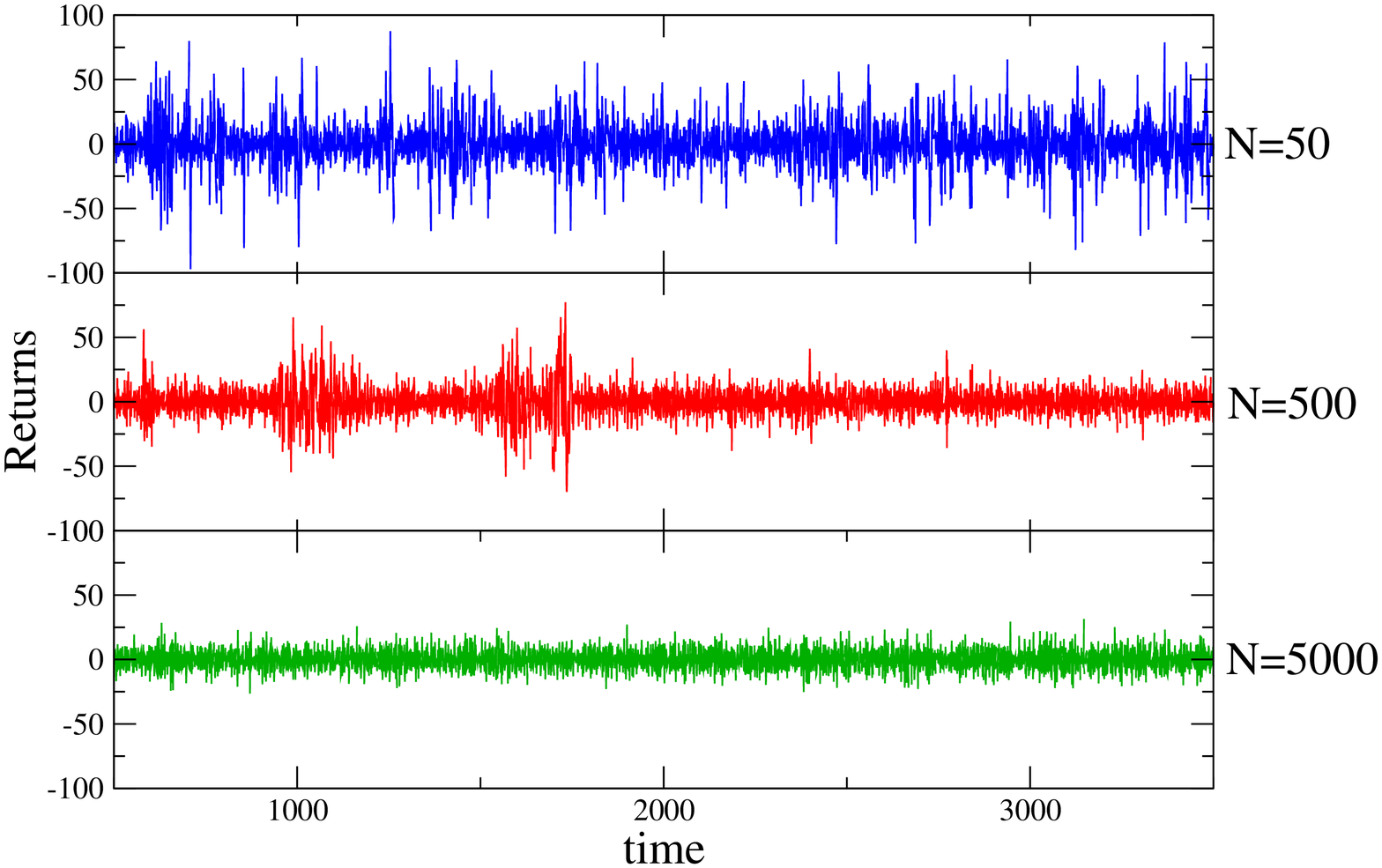}}
\caption{Dynamical evolution of the agents strategies (a) for the model discussed in Fig.~\ref{fig:1} and the corresponding fluctuations of returns (b). One can see that the intermittent behavior of returns (which leads to the stylized facts) is obtained only for $N=500$. This is due to the fact that for large $N$ (5000) the system is too stable in the fundamentalist state, while for smaller values of $N$ (50) it fluctuates too rapidly. The value of $N$ is in fact crucial for the transitions between metastable states for the agents strategies. It is clear therefore that the intermittent region corresponds to finite size effects which disappear for $N$ very large.}
\label{fig:2}
\end{figure}
For small price fluctuations the use of linear or log returns are essentially equivalent and in this paper
we prefer the linear description because it permits an analytical treatment of various properties.
A crucial element in the dynamics is the fact that, at each time step, an agent can change its strategy from chartist to fundamentalist or viceversa\cite{kirman,marsili,gallegati,rcont3}. The probability for this change depends on two basic elements,  an herding effect and a function which contains information on the price. We also add a constant probability $K$ which corresponds to the possibility of neglecting the opinion of the  other agents if the price signal is strong enough. Without such a term the dynamics would be  intrinsically unstable toward  absorbing states corresponding to all fundamentalists or all chartists. 
The rate equation for the transition from chartists to fundamentalists is  given by
\begin{eqnarray}
P_{cf}\propto(K+\frac{N_f}{N})\exp(\gamma |p_f-p|)
\label{eq:5}
\end{eqnarray}
and the rate from fundamentalists to chartists is
\begin{eqnarray}
P_{cf}\propto (K+\frac{N_c}{N})\exp(b \frac{ |p_M-p|}{M-1})
\label{eq:6}
\end{eqnarray}
The dynamics of the strategies of the agents is a key point in the problem. Neglecting for the moment the exponential price terms, the asymptotic stable distribution $P_e(x)$ for the two classes is given by~\cite{kirman}:
\begin{equation}
P_e\propto x^{\epsilon -1}(1-x)^{\epsilon-1}
\label{eq:7}
\end{equation}
where $x=N_c/N$ and $\epsilon=KN$.
The characteristic duration $T_0$ of a metastable state can also be explicitly derived.
The case discussed corresponds to a perfect symmetry between the two classes of agents. In reality the two classes are not symmetric because fundamentalists usually correspond to institutional investors and have (on average) a major role in the market~\cite{mantegna,bouchaud,alfi3}. 
This can be reproduced by introducing an asymmetry $(\delta)$ in the rates
\begin{eqnarray}
P_{cf}=B(1+\delta)(K+\frac{N_f}{N})\\
P_{fc}=B(1-\delta)(K+\frac{N_c}N)
\label{eq:8}
\nonumber
\end{eqnarray}
where in this case $K= r/N$ in order to be able to vary $N$ without affecting the
parameters space of the model.
In this case the asymptotic stable distribution also reflects the asymmetry
and an approximate analytical expression can be obtain~\cite{alfi1}
\begin{eqnarray}
P_{eq}\sim x^{r(1-\delta)-1}(1-x)^{r(1+\delta)-1}\exp{(-2\delta \nu N)}.
\label{eq:9}
\end{eqnarray}
The basic behavior is that the system is mostly in a state with predominance of fundamentalists and it occasionally makes a transition to a state dominated by chartists. Both these metastable states have characteristic duration times which can be estimated~\cite{alfi1}. These characteristic times depend crucially on the number of agents $N$ and for very large $N$ they diverge. This means that a change of state is rather easy for small $N$ and it becomes progressively harder for increasing values of $N$.
In Fig.~\ref{fig:1} we show the stable distribution for different values of $N$ and in Fig.~\ref{fig:2} we report the corresponding time series. The intermittent behavior made of bursts in the number of chartists occurs only for an intermediate value of $N$ (500). This will lead to the SF which therefore correspond to finite size effects in the sense that they disappear in the large $N$ limit. This result has important conceptual and practical implications. In fact, in such a situation, the effective exponent usually adopted to characterize the SF cannot be expected to be universal. This means that specific effects in the real data and in the model can affect their values, which should not be considered as fundamental test for the models.
In Fig.~\ref{fig:3} we report the Stylized Facts corresponding to the intermittent case $(N=500)$. 
The fat tails are clearly present and so is the volatility clustering. In its simplest version the model leads to a single characteristic time for the agents fluctuations. In this case we have effectively a single stochastic agent and the volatility correlation decays exponentially. However, if we introduce an heterogeneity in the agents, for example by considering different time horizons in the moving averages $(M)$ a longer relaxation can be obtained as shown in the right insert and in the main figure. This can be fitted by a power law exponent but its value is not universal and it depends on the parameters used. The interesting side of this situation is that one should not consider as a problem the fact that different stocks, different time scales and different markets seem to lead to apparent exponents which are not identical. Our conclusion is that
this is not a limitation of the data but an intrinsic property of the dynamics.
\begin{figure}[htbp]
\centering
\includegraphics[scale=0.4]{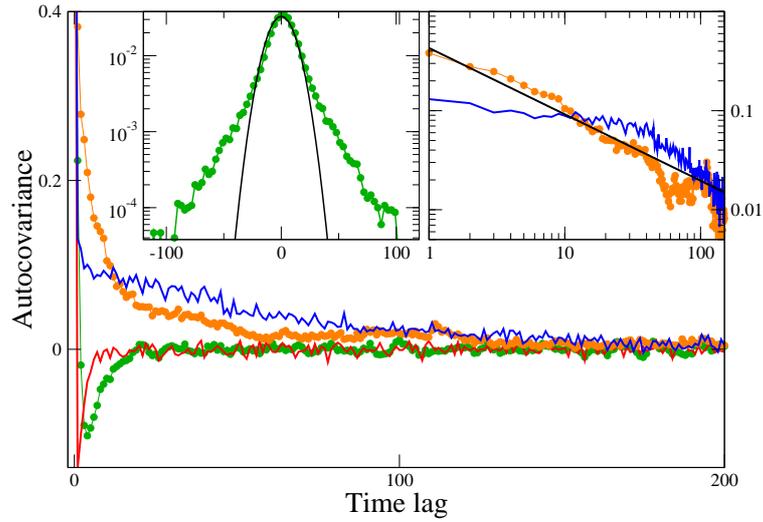}
\caption{A summary of the Stylized Facts obtained from the model. The left insert shows the fat tails for the returns. The right insert shows the volatility clustering. The blue line corresponds to agents with a single characteristic time for the transition between the two strategies. In this case we have effectively a single stochastic agent and the decay of the volatility correlations is exponential. If one consider the heterogeneous situation in which agents consider different time horizons in their mobile averages (values of $M$) one can see that a more complex behavior is obtained which resembles a power law. Its effective exponent, however, is not universal and can be tuned by the specific parameters considered. In the main part of the figure we show the auto covariance (in linear scale) for the volatility and for the returns. One can see that returns show a small negative correlation at short times while the correlation of the variance extends to much longer times.}
\label{fig:3}
\end{figure}

\section{Nonstationarity and the Microscopic Origin of Stylized Facts}

The simplicity of the model permits a detailed interpretation of the origin and nature of the stylized facts. The long time correlation of volatility and its behavior in terms of bursts depends essentially on two elements. The first is that there is a certain probability to have an appreciable number of chartists and this situation tends to destabilize the price for the duration of this metastable state. 
This can occur without a specific reason like in the sandpile model and in self-organized critical systems~\cite{perbak}.
A price fluctuation tends to increase the chartists and they will bet on an even larger fluctuation. This will increase the number of chartists etc. This leads to a multiplicative effect in which the variance at a time $t+1$ depends essentially on the variance at time $t$ and the price change.
These are essentially the elements which motivated the introduction of the phenomenological GARCH model~\cite{mantegna,bouchaud}, schematically corresponding to the relation below.
\begin{equation}
\sigma(t+1)=f(\sigma(t);\Delta p(t))
\label{eq:10}
\end{equation}
The difference, however, is that now we can give a microscopic interpretation of this phenomenological ansatz in terms of the agent and price dynamics.
\\
In the model the arbitrage condition is not explicitly implemented, nevertheless the correlations of the returns appears to be negligible (Fig.~3). This can be understood by considering that the simple arguments above about the variance are based on the simple phenomenon that action leads to more action, independent of the direction. For the returns the situation is different because the specific direction of the price depends on many more additional elements like the specific distribution of agents, their strategies and the specific dynamics of $p(t)$. All these elements are much less correlated with respect to the variance and we can expect a much more complex relation as indicated below.
\begin{equation}
\Delta p(t+1)=f(\sigma; N_c; N_f; p_M(t); p_f)
\label{eq:11}
\end{equation}
This provides a qualitative understanding of the absence of correlations in returns even if this arbitrage property is not explicitly implemented in the model.We expect these conditions to have a general validity beyond this specific model.
\begin{figure}[htbp]
\centering
\includegraphics[scale=0.4]{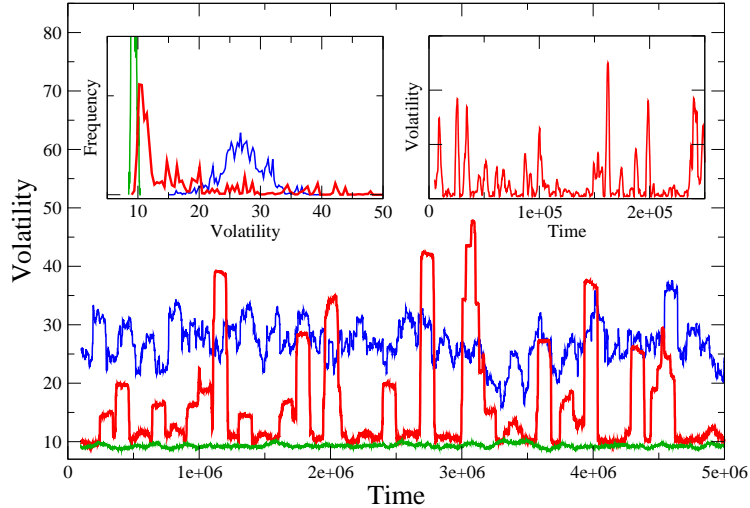}
\caption{The main figure shows the time series of the volatility corresponding to $N=50$ (green), $N=500$ (red) and $N=5000$ (blue). In the left insert we show the frequency distribution of the volatility corresponding to the three cases. In the right insert we show the volatility of the price of a real stock. One can see that its intermittent behavior is rather similar to the one we obtain for the case $N=500$ (red line).}
\label{fig:4}
\end{figure}

\section{Self-Organization of Stylized Facts}

Finally we can consider the problem of the self-organization of the system in the intermittent state. In fact, also in our model, the SF are obtained only for specific values of the parameters and a basic question is to understand why the systems chooses to stabilize in this special situation. The key point in this respect is to consider that the number of agents $N$ is not necessarily fixed a priori and constant in the dynamics of the system. Agents are usually attracted to a certain market if there is some action in it. A price with very small fluctuations is often considered as not interesting for trading. So the basic idea is that the agents enter the market and act on it only if the price fluctuations are above a certain threshold. Considering the fluctuations over a time interval $T$ given by the expression:
\begin{equation}
\sigma(t, T)=\frac{1}{T-1}\sum_{i=t}^{t-T}(p_i-\bar p)^2
\label{eq:12}
\end{equation}
an agent will enter or exit from this market under the conditions:
\begin{eqnarray}
\sigma(t, T)>\Theta_{in}\\
\sigma(t, T)<\Theta_{out}
\label{eq:13:14}
\end{eqnarray}
where $\Theta_{in}$ and $\Theta_{out}$ are suitable thresholds.
\begin{figure}[htbp]
\centering
\includegraphics[scale=0.4]{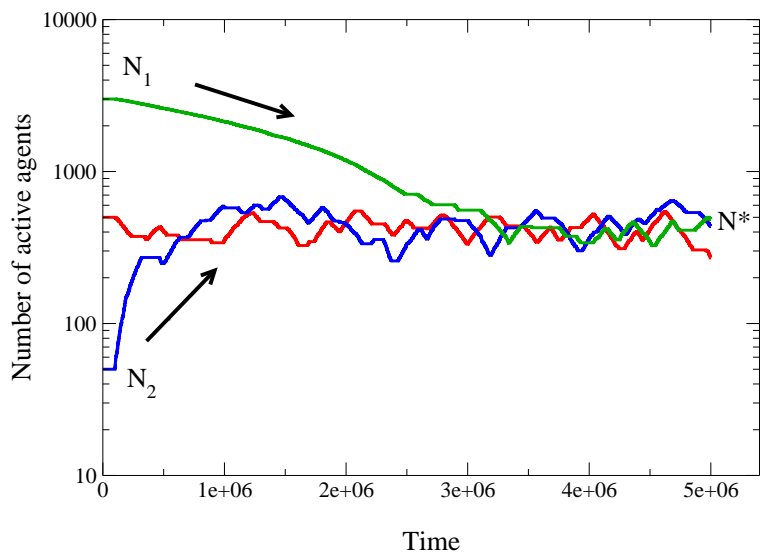}
\caption{Example of the self-organization of the system in the quasi-critical state corresponding to the intermittent behavior and the stylized facts. The key point for the self-organization is the introduction of a threshold for an agent to be active which depends on the price fluctuations. If we start with a large number of agents ($N=5000$) the system is rather stable in the F state and price fluctuations are very small. This will induce agents to become inactive and $N$ decreases (green line). If we start from a small number ($N=50$) fluctuations are large and this will attract agents to enter this market (blue line). The critical number of agents ($N=500$ for our choice of parameters) remains essentially stable (red line). This example of the self-organization of the market towards an  intermittent state emphasizes the importance of the nonstationarity properties of the system.}
\label{fig:5}
\end{figure}

In Fig.~\ref{fig:4}  we show these fluctuations for different values of $N$. One can see that for small values of $N$ ($N=50$) the fluctuations are rather large, while for large values $(N=5000)$ they are very small. Only for the intermediate case $N=500$ one has the intermittent behavior which leads to the SF and resembles closely the behavior of real prices shown in the right insert.
We can now consider the dynamics of a system in which N can vary according to the thresholds we have discussed. In Fig.~\ref{fig:5} we can in fact observe the  phenomenon of self-organization. Starting from a small value of $N$ $(N=50)$ the large fluctuations will attract more agents and $N$ increases. Starting instead from a large value $(N=5000)$ the opposite happens and $N$ decreases. For $N=500$ we have a relatively stable situation. The self-organization to the quasi-critical state with SF corresponds therefore to the fact that this is the attractive fixed point for the dynamics of the agents  related to their threshold strategy.
The fact that this occurs in our model for $N=500$ depends on the specific parameters we have adopted but clearly the phenomenon of self-organization is completely general and robust.
In principle, by changing the parameters, one could have SF stabilized at different values of $N$.

\section{Conclusions}
In summary the model we have introduced was motivated by the objective of having the maximum mathematical simplification that is still able to reproduce the stylized facts. This has permitted a detailed interpretation of their nature and origin of SF and it is a crucial point in order to consider the phenomenon of the self-organization. It is clear that the model can then be modified in many ways in order to consider more realistic situations for the description of the market. These more realistic cases, however, can now be addressed in a systematic way.
In the following lines we draw the main conclusions of our paper.
\\
The SF correspond to finite size effects and not to critical exponents even though they
can occur at different time scales. This result has important conceptual and practical implications.
\\
The specific origin of fat tails and volatility clustering is identified in a multiplicative process which can occur spontaneously as in the sandpile model~\cite{perbak}.
\\
The absence of correlations in the returns can be understood
along similar lines.
\\
Non-stationarity in the number of agents and in their action
on the market is an essential point.
\\
The variability of the action is a new element for the self-organization of the system towards
the quasi-critical state.
\\
It should be possible to test many of these conclusions with respect to the general phenomenology
of real markets.

\end{document}